\batchmode 

\documentclass{article}

\begin{document}

\large

\title{ Feynman Equation in Hamiltonian 
Quantum Field Theory} 

\author{ Alexander Dynin \\ 
Ohio State University, Columbus, Ohio 43210, USA \\ E-mail: 
dynin@math.ohio-state.edu}

\date{}

\maketitle

\begin{abstract}
Functional Schr\"{o}dinger equations for interacting fields
 are solved via rigorous non-perturbative  Feynman type integrals. 
 \end{abstract}



\large

\medskip
\begin{center}
\textsc{to i.e.segal, in memoriam}
\end{center}

\section{Introduction.}
Semantically, quantum field theory means either a theory of
quantum fields, or  a quantum theory of fields.
Mathematically, it describes either  a classical evolution of operator
fields, or  an operator evolution of classical fields.

\smallskip
The \emph{ Feynman  equation in  quantum theory of fields}  is
$$
\langle \phi'',t'' |\phi',t'\rangle = 
\sum_{\phi'}^{\phi''} e^{i\mathcal{A}_{t'}^{t''}(\phi)}.
$$
(We assume the Planck constant $\hbar=1$.)

The left hand side of the equation is the \emph{probability amplitude} 
of a quantum transition between classical  fields $\phi'(x)$ and 
$\phi''(x)$ on the euclidean space $\mathbf{R}^{d}$ at times $t'$ 
and $t''$. 

The right hand side of the equation is a \emph{sum over classical histories} 
$\phi = \phi(x,t)$ from $t'$ to $t''$
on the Minkowski space-time $\mathbf{R}^{d,1}$.

 The \emph{action functional} $\mathcal{A}_{t'}^{t''}$ is
$$
\mathcal{A}_{t'}^{t''}(\phi) = \int_{t'}^{t''}dt 
\int_{\mathbf{R^{d}}}dx\,l(\phi,\bigtriangledown\phi,\dot{\phi}),
$$
where the Lagrangean density $l$ is a Lorentz 
invariant real function of $\phi$ and its space and time  derivatives
 $\bigtriangledown\phi,\dot{\phi}$.

\smallskip

A short-hand notation for the ``sum'' is the 1948 \emph{Lagrangean Feynman 
integral} 
$$
\int_{\phi'}^{\phi''} \mathcal{D}(\phi)\, 
e^{i\mathcal{A}_{t'}^{t''}(\phi)}.
$$
This compact notation suggests the \emph{standard algorithms of the elementary 
integral calculus}, including repeated integration, integration by 
parts,  substitution rule, WKB approximation, Gaussian integrals and so 
on.

Another form of the Feynman integral is a \emph{ Hamiltonian Feynman 
integral} (R.Feynman [9] in 1951, W.Tobocman [19] in 1956)
$$
\int_{\phi'}^{\phi''} D(\phi,\pi)\, 
e^{i\mathcal{A}_{t'}^{t''}(\phi,\pi)},
$$
where $\pi = \partial l/\partial\phi_{t}$ and  
$\mathcal{A}_{t'}^{t''}(\phi,\pi) = \int_{t'}^{t``}dt 
\int_{\mathbf{R}^{d}}dx\, [\pi\dot{\phi}- 
h(\phi,\pi)]$
 with the \emph{ Hamiltonian function}
$h(\phi,\pi) = \pi\dot{\phi}- 
l(\phi,\bigtriangledown\phi,\pi).$

\smallskip 
In 1960 J.Klauder [12] introduced 
the \emph{Feynman integral over the coherent state histories} 
with $h$ being the Wick symbol of the quantum evolution.

\smallskip 
In 1973 E.Lieb [15] 
modified the Klauder construction using  the anti-Wick symbol
of the quantum evolution. 

\smallskip
Until now, in spite of their fundamental 
importance in quantum field theory, 
these Feynman integrals have been largely unjustified
in rigorous mathematical terms. 

\smallskip
According to canonical formalism, the quantum evolution in the left hand side of 
the  Feynman equation is defined by a self-adjoint Hamiltonian operator 
$H$. However, until now even its domain  has been largely problematic.

\smallskip
\emph{In this paper, for a wide variety of interaction Hamiltonians, 
we define rigorously  both sides of the Feynman equation  and show that
they are equal under  appropriate conditions.}

In particular, we establish a rigorous  equivalence  of the 
corresponding canonical and path integral quantizations in such cases.

 \smallskip
 The construction of the Feynman type functional integral is fairly new.
  As in the Feynman original approach, it is of
 sequential type, but the Feynman-Tobocman semi-classical 
 postulate for short time  propagators is modified as in [8]. 
 
 \smallskip 
 We use the Feynman-type integral 
 to solve the corresponding functional temporal Schr\"{o}dinger 
 equation via a limit of  multiple  functional 
 integrals over the infinite-dimensional phase space.
  
\emph{Because the solution is non-perturbative, all renormalization problems 
are circumvented.}

\smallskip
The main results are in the section 4. They have been partially presented at the 
Conference on \emph{Feynman Integrals and Related Topics}, July, 1999,
Seoul, Korea, and at the Special Session on \emph{The Feynman Integral 
with Applications} of the  Annual Meeting of the American Mathematical
Society, Washington, D.C.,January, 2000.

\smallskip
This paper is an independent sequel of [8].

\section{ Review of Segal boson systems.}

\subsection{Segal boson system (cf.[2]).}

      The \emph{Segal boson system} $(\mathcal{F},\widehat{\cdot},\Omega, H_{0})$
over a phase space $\mathcal{H}$ is a unversal model for  concrete  
free boson  fields of positive mass. According to I.Segal, this is 
a universal free boson  field [2].

\smallskip
The \emph{phase space} $\mathcal{H}$ is a complex separable Hilbert space
with a hermitean sesquilinear form $\langle\cdot|\cdot\rangle$. 
By physicists convention, the form  is antilinear on the left.

As a phase space, $\mathcal{H}$ is a 
 symplectic vector space with the symlectic form 
$\Im\langle\psi_{1} | \psi_{2}\rangle$,
 the imaginary part of the Hermitean product on $\mathcal{H}$.

\smallskip
The four constituents $(\mathcal{F},\widehat{\cdot},\Omega, H_{0})$
are defined axiomatically as folows.
\begin{itemize}
\item The \emph{abstract Fock space} $\mathcal{F}$ is
a complex Hilbert space of quantum states $\Psi$.
\item The \emph{Heisenberg canonical commutation
 relation $\widehat{\cdot}$ (or CCR)}  is  a continuous
$\mathcal{R}$-linear mapping  of  phases
 $\psi\in\mathcal{H}$ to self-adjoint operators $\widehat{\psi}$
 on  $\mathcal{F}$ satisfying 
$$
 \widehat{\psi}_{1}\widehat{\psi}_{2} - \widehat{\psi}_{2}\widehat{\psi}_{1}
 = -i\Im\langle\psi_{1}|\psi_{2}\rangle\mathbf{1}.
$$
\item  The  \emph{vacuum} $\Omega\in\mathcal{F}$ is a fixed
fiducial quantum  state, i.e.,  the linear span of
 $\exp(i\widehat{\psi})\Omega,\  \psi\in \mathcal{H}$,  
 is dense in $\mathcal{F}$.
 \item The \emph{free Hamiltonian operator} $H_{0}$ 
 is a non-negative non-zero self-adjoint operator on $\mathcal{F}$
 such that $H_{0}\Omega = 0$ and
$$
H_{0}\widehat{\psi}-\widehat{\psi}H_{0}=i\widehat{\psi}.
$$
\end{itemize}
 The continuity, linearity and the commutator relations for $\widehat{\psi}$
are understood in terms of unitary operators $\exp(i\widehat{\psi})$
(cf.[2]). 

\smallskip
In spite of  uncountably many  unitary non-equivalent CCR, we have 
the following  fundamental Segal's theorem [2]:

\begin{itemize}
\item \emph{A Segal boson system 
$(\mathcal{F}, \widehat{\cdot}, \Omega, H_{0})$
 over a  phase space $\mathcal{H}$ is unique up to unitary eqivalence.}
 
 \item \emph{For every self-adjoint operator $a$ on $\mathcal{H}$ there is a 
unique self-adjoint operator $\widehat{a}$ on $\mathcal{F}$ such that 
for all  $\psi\in\mathcal{H}$}
 $$
\widehat{a}\widehat{\psi}-\widehat{\psi}\widehat{a} = \widehat{a\psi}
$$
In particular, $\widehat{\mathbf{1}} = H_{0}$.

\item \emph{Moreover, $\widehat{a} \geq 0$ if $a \geq 0$}. 
 
\end{itemize}
In view of  Segal's fundamental theorem, 
the Segal boson system is defined by the  dimension of its phase space.

\smallskip
The Schr\"{o}dinger formulation of the  quantum mechanical
harmonic oscillator is the Segal system over a finite-dimensional 
phase space. It is known 
as a \emph{first quantization} of the classical harmonic oscillator.
Its Fock space is  a single particle space.

\smallskip
If one takes the infinite dimensional  Fock space of
the first quantization as a new phase space, the corresponding 
Segal system is the \emph{second quantization} of the
classical harmonic oscillator. Its free Hamiltonian $H_{0}$
is the \emph{number operator}.

\smallskip
The Fock-Cook tensor representation of the CCR for a free
 relativistic boson system of positive mass (cf.[5]) 
is the  second quantized Segal system.

\smallskip
Along with the Fock-Cook tensor representation of the Segal system, 
two Gaussian representations are most important (cf.[2]).

\subsection{ Real Gaussian representation.}

A \emph{conjugation} 
on the phase space $\mathcal{H}$ is an 
antilinear isometric involution $\psi\rightarrow\psi^{*}$.
The invariant phases $\phi = \phi^{*}$ form the 
corresponding real part $\Re^{*}\mathcal{H}$ 
of $\mathcal{H}$.

Let $D[\phi]$ be the functional measure on $\Re^{*}\mathcal{H}$
 defined as a weak inductive limit of the euclidean measures on 
real finite-dimensional subspaces with the euclidean scalar products 
$\Re\pi\langle\phi'' | \phi'\rangle$ (cf., e.g.,  
[4], where the functional  measure is defined
 on a pre-hilbert  space and shown to satisfy  
the  standard rules of elementary integral calculus). 

\smallskip
The functional Gaussian measure 
$e^{-\langle\phi | \phi\rangle}D[\phi]$
is  the corresponding weak inductive limit of  Gaussian
 measures on  real finite-dimensional subspaces of $\Re^{*}\mathcal{H}$
 (cf. [4]).

\smallskip 
The Fock space $\mathcal{F}$ of the real Gaussian representation 
is the Hilbert 
space completion $\mathcal{L}^{2}(\Re^{*}\mathcal{H},
e^{-\langle\phi | \phi\rangle}D[\phi])$
of the complex span of the real-analytic polynomials $\Psi(\psi)$
on $\Re^{*}\mathcal{H}$.

For real* phases $\phi=\phi^{*}$ the CCR representation is 
$$
(\widehat{\phi}\Psi)(\phi') = \langle\phi | \phi\rangle \Psi(\phi'),
$$
where  $d_{\psi}$ is the  functional derivative 
in the direction of $\psi$.

For imaginary* phases $\psi=i\phi$ the CCR representation is 
$$
(\widehat{i\phi}\Psi)(\phi') = \frac{1}{i}(d_{\phi}\Psi)(\phi)
+\frac{1}{i}\langle\phi | \phi'\rangle\Psi(\phi').
$$
 
 \smallskip
 The vacuum vector $\Omega$ is the Gaussian vector 
 $\exp(-\Re\langle\phi | \phi\rangle)$.

\smallskip
The free Hamiltonian $H_{0}$ is $d^{\dagger}d$, where 
$d$ is the   functional differential and $d^{\dagger}$ 
is its Hermitian adjoint. 

\subsection{Complex Gaussian representation.}

The Fock space of the complex Gaussian representation is 
the closure  $\mathcal{B}^{2}(\mathcal{H})$ in the  Hilbert 
space $\mathcal{L}^{2}(\mathcal{H},
e^{-\langle\psi | \psi\rangle}D[\psi])$
of the complex span of the \emph{antiholomorphic polynomials} 
$\Psi(\psi)$ on $\mathcal{H}$.  
The functional measure $D[\psi]$ is defined as the inductive limit of 
the euclidean measures on  finite-dimensional \emph{complex}  subspaces
 with the euclidean scalar products  $\pi\langle\psi'' | \psi'\rangle$.

\smallskip
The CCR representation is 
$$
(\widehat{\psi}\Psi)(\psi') = 
\frac{1}{\sqrt{2}}\left[\overline{\partial}_{\psi}
\Psi(\psi') + \langle\psi' | \psi\rangle)\Psi(\psi')\right],
$$
where $\overline{\partial}_{\psi}$ is  the anti-complex functional derivative.
 
\smallskip
The vacuum vector $\Omega$ is the constant $1$.

\smallskip
The free Hamiltonian $H_{0}$ is $\overline{\partial}^{\dagger}
\overline{\partial}$, 
where $\overline{\partial}$ is the (unbounded)  functional 
anti-complex differential on 
$\mathcal{B}^{2}(\mathcal{H})$ and $\overline{\partial}^{\dagger}$ 
is its Hermitian adjoint.

\subsection{ Hilbert scales (cf.[16] and [2]).}

A self-adjoint operator $s \geq \mathbf{1}$ on $\mathcal{H}$ is called
a \emph{scaling operator}.

For $\rho\geq 0$ define the Hilbert spaces
$$
\mathcal{H}_{\rho} = \{\psi \in \mathcal{H}: 
||\psi||_{\rho} = ||s^{\rho/2}\psi||<\infty\},
$$
and the Hilbert spaces
$\mathcal{H}_{-\rho}$ which are the  completion of $\mathcal{H}$
relative to the norm 
$||\psi||_{\rho} = ||(\mathbf{1} + s)^{-\rho/2}\psi||$.  

\smallskip 

The family of ($\mathcal{H}_{\rho},\  \rho\in\mathbf{R}$) 
form a \emph{Hilbert scale}: $\mathcal{H}_{\rho +1}$
are densely and continuously imbedded in $\mathcal{H}_{\rho}$. 

\smallskip
The topological intersection
$\mathcal{H}_{\infty} = \cap \mathcal{H}_{\rho}$
is the \emph{core} of the scale. The core  is naturally 
a Frechet space. 

\smallskip
A Hilbert scale is \emph{Hilbert-Schmidt}
 if the inverse of the scaling operator is a Hilbert-Schmidt operator.
The core of a Hilbert-Schmidt scale is nuclear.

\smallskip
The spaces $\mathcal{H}_{\rho}$ and $\mathcal{H}_{-\rho}$
are anti-dual relative to the basic hermitian form $\langle\cdot | 
\cdot\rangle$, and so are the core and the topological union 
$\mathcal{H}_{-\infty} = \cup \mathcal{H}_{-\rho}$.

\smallskip
The Hilbert scale construction is applicable to the  state space
$\mathcal{F}$ with the scaling operator $\mathbf{1}+\widehat{s}$.  
This gives a Hilbert scale $(\mathcal{F}_{\rho})$ with the 
 core $\mathcal{F}_{\infty}$ and its anti-dual $\mathcal{F}_{-\infty}$.

 \smallskip
\emph{Henceforth, we deal with infinite-dimensional 
phase space $\mathcal{H}$ only.} In its first quntized realization we choose 
the one-dimensional  harmonic oscillator to be  the scaling operator $s$. 
Now the core $\mathcal{H}_{\infty}$ is nuclear (though the core
$\mathcal{F}_{\infty}$ is not), and $\Omega\in\\mathcal{F}_{\infty}$. 

Moreover, the vacuum vector 
$\Omega\in\mathcal{F}_{-\infty}$.

\smallskip
In the definitions of the functional measures $D[\phi]$ and
$D[\psi]$ it is possible 
to choose the finite-dimensional subspaces from the core 
$\Re^{*}\mathcal{H}_{\infty}$. 
Thus the integration over $\Re^{*}\mathcal{H}$ and $\mathcal{H}$ 
coinside with the integration over $\Re^{*}\mathcal{H}_{\infty}$ 
and $\mathcal{H}_{\infty}$.
 
\smallskip
 Since the core $\mathcal{H}_{\infty}$
 is nuclear,  the Minlos theorem (cf.[10]) states that the 
integration relative to the functional Gaussian measures is equivalent to  
the integration relative to the Radon  
 Gaussian measure on $\mathcal{H}_{\infty}$.

\subsection{Coordinate representations.}
 
 Consider a topological subset
  $X\subset\mathcal{H}_{-\infty}$ with  a Borel or a functional
measure $\mu(\chi)$ on $X$. 
 
\smallskip
By definition, $(X, \mu)$ is a\emph{Plancherel basis} 
in $\mathcal{H}$ if 
$\int_{X}d\mu(\chi)|\chi\rangle\langle\chi|$
is an orthogonal resolution of the identity 
$\mathbf{1}$ on $\mathcal{H}$. 
This means that
if $\psi(\chi) = \langle\chi |
 \psi\rangle$ for $\psi\in\mathcal{H}_{\infty}$ then
$\psi  = \int_{X}d\mu(\chi)\psi(\chi)$ and
$\langle\psi'' | \psi'\rangle = 
\int_{X}d\mu(\chi)\overline{\psi''(\chi)}\psi'(\chi).$

\smallskip
By extension, these two equations hold for all
$\psi, \psi', \psi''\in \mathcal{H}$. Then $\psi(\chi)$ are defined only 
for almost all $\chi$. 

\smallskip
By a version of the spectral Gelfand-Kostuchenko theorem,
\emph{a finite family of  commuting self-adjoint operators on 
$\mathcal{H}$ has a common Plancherel eigenbase $X$ because
$\mathcal{H}_{\infty}$ is nuclear}. 

\smallskip 
Examples: 

1) $\mathcal{H}$ is a functional space 
 $\mathcal{L}^{2}(\mathbf{R}^{d},\gamma(x)dx)$ with 
 $\gamma\in\mathcal{L}^{1}(\mathbf{R}^{d})$.
The commuting self-adjoint operators
are multiplications with coordinate functions. Then 
$(\mathbf{R}^{d},\gamma(x)dx)$ is a Plancherel basis. 

\smallskip
2) Let $X$ be the set of the eigenvectors of the scaling operator $s$ and 
$\nu$ be the counting measure on $X$. Then $(X,\nu)$ is a Plancherel basis.
In the Plancherel expansion of $\mathcal{H}$ the scaling 
operator $s$ acts on $\psi$ as the multiplication $\lambda(\chi)\psi(\chi)$
 with the eigenvalues $\lambda(\chi)$.

\smallskip
3) Suppose a conjugation  on  $\mathcal{H}$ commutes with $s$.
Then it defines a conjugation on all $\mathcal{H}_{\rho}$, 
and  we get real Hilbert scales $\Re^{*}\mathcal{H}_{\rho}$.

Then in the associated Plancherel 
expansion $\mathcal{H} = \mathcal{L}^{2}(X,\mu)$ the  
conjugation becomes the usual complex conjugation. 

If $\mathcal{H}=\mathcal{L}^{2}(R^{d},\gamma(x)dx)$, then 
the real phases represent  \emph{classical fields}.

\subsection{Annihilation and Creation operators 
 (cf.[1]).}

For every $\psi$ in $\mathcal{H}$ define a closed 
\emph{annihilation operator} on $\mathcal{F}$
$$
A(\psi) = [\widehat{\psi} + i(\widehat{i\psi})]/\sqrt{2}.
$$
It is \emph{antilinear} in the parameter $\psi$ and annihilates 
the vacuum $\Omega$. 

\smallskip
The annihilation operators  are continuous 
 on $\mathcal{F}_{\infty}$ and commute.
 Moreover  they are strongly continious in the parameter $\psi$
 relative to  the topology of $\mathcal{H}_{-\infty}$. Then  by 
continuous extension in  $\psi$, they are defined for every 
$\psi\in \mathcal{H}_{-\infty}$ 
as commuting continuous operators on $\mathcal{F}_{\infty}$. 

The sesquilinear map $(\psi, \Psi) \rightarrow A(\psi)\Psi$ is 
jointly continuous from $\mathcal{H}_{-\infty}\times\mathcal{F}_{\infty}$
to $\mathcal{F}_{\infty}$. 

In particular, the annihilation operators $A(\chi)$ are well
defined on $\mathcal{F}_{\infty}$ and parametrically continuous on $X$

\smallskip

The \emph{creation} operators $C(\psi),\ \psi\in \mathcal{H}_{-\infty}$, 
are  the  adjoints of $A(\psi)$ relative to the hermitian form 
$\langle\cdot | \cdot\rangle$. They are continuous 
operators on $\mathcal{F}_{-\infty}$ and commute. The creation 
operators $C(\psi)$ are linear in $\psi\in\mathcal{H}_{-\infty}$.

\smallskip 
In the complex Gaussian representation $A(\psi) = \overline{\partial}_{\psi}$ 
and $C(\psi)$ is the multiplication with $\langle\psi | \cdot\rangle$. 

\smallskip
In  $(X,\mu)$-coordinates
 $$
A(\psi) = \int_{X}d\mu(\chi)\overline{\psi(\chi)}A(\chi),\quad
C(\psi) = \int_{X}d\mu(\chi)\psi(\chi)C(\chi).
$$ 
 in $\mathcal{F}_{\infty}$ and $\mathcal{F}_{-\infty}$, correspondingly.

\smallskip
The \emph{coherent states} are 
$e^{\psi}=\sum_{n=0}^{\infty}\frac{1}{n!}C(\psi)^{n}\Omega,\ 
\psi\in\mathcal{F}$. Thus $\langle e^{\psi''} | e^{\psi'} \rangle = 
e^{\langle\psi'' | \psi'\rangle}$.

\smallskip
In the complex Gaussian representation, 
the \emph{coherent}  states are
 $e^{\psi}(\psi') = e^{\langle\psi' | \psi\rangle}$.

\smallskip 
A coherent state $e^{\psi}$ belongs to $\mathcal{F}_{\infty}$
if and only if $\psi\in\mathcal{H}_{\infty}$. 

\smallskip  
Set  $\Psi(\psi) = \langle \Psi | e^{\psi}\rangle$
for  $\Psi\in \mathcal{F}$. Then
$$
\langle\Psi'' | \Psi'\rangle  = \int_{\mathcal{H}}D[\psi]
 e^{-\langle \psi| \psi\rangle} \overline{\Psi''(\psi)}\Psi'(\psi).
$$
Thus ($\mathcal{H}, e^{-\langle \psi| \psi\rangle}D[\psi]$)
is  a Plancherel basis in $\mathcal{B}^{2}(\mathcal{H})$.

\section{Revision of the Lascar infinite-dimensional 
pseudodifferential operators (cf. [14]).}

\subsection{Wick and Berezin symbols.}

Consider a continuous linear operator $Q$ from $\mathcal{F}_{\infty}$
 to $\mathcal{F}_{-\infty}$. Among its various integral kernels 
 we have the \emph{coherent state matrix element}
$\langle e^{\psi''}|Q|e^{\psi'}\rangle$: 
$$
(Q\Psi)(\psi'') = \int D[\psi']\,e^{\langle\psi''-\psi' | \psi'\rangle}
\langle e^{\psi''}|Q|e^{\psi'}\rangle. 
$$
As an entire function 
of $\psi', \psi''$ on  
$\overline{\mathcal{H}_{\infty}}\times\mathcal{H}_{\infty}$, the
coherent state matrix element is completely defined by its  
restriction $\langle e^{\psi}|Q|e^{\psi}\rangle$ to the real diagonal.   

\smallskip 
The \emph{Wick (or normal) symbol} $Q^{w}$ of $Q$ is
$$
Q^{w}(\psi) = e^{-\langle\psi|\psi\rangle}\langle e^{\psi}|Q|e^{\psi}\rangle.
$$
Suppose $\mathcal{F}_{\infty}$ is invariant under $Q_{1}$ and $Q_{2}$. 
Then the operator  product $Q_{2}Q_{1}$ is well defined on
$\mathcal{F}_{\infty}$ with the Wick symbol
$$ 
(Q_{2}Q_{1})^{w}(\psi) = \int D[\psi']\,e^{-\langle\psi - \psi' | \psi'\rangle}
Q_{2}(\psi,\psi') Q_{1}(\psi',\psi).
$$
The Wick symbol of the adjoint operator $Q^{\dagger}$ is
 $\overline{Q^{w}(\psi)}$.
Thus the operator $Q$ is symmetric on $\mathcal{F}$ if and only if
its Wick symbol is real. 
 
\smallskip
If an operator $Q$ has a \emph{Toeplitz integral kernel}:
$$
(Q\Psi)(\psi'') = \int D[\psi']\,
 e^{\langle\psi''-\psi'|\psi'\rangle}Q^{b}(\psi')\Psi(\psi'),
$$
with $Q^{b}\in\mathcal{F}_{\infty}$
then $Q^{b}(\psi)$ is the \emph{Berezin symbol} of $Q$.
It is  uniquely defined by $Q$.

\smallskip
If $Q^{b}$ exists then the Berezin symbol of the adjoint $Q^{\dagger}$
exists and equals to $\overline{Q^{b}(\psi)}$.

Thus the operator $Q$ is symmetric on $\mathcal{F}$ if and only if
$Q^{b}(\psi)$ is real.

\smallskip
The decisive advantage of the Berezin symbol is  that the numerical range
$\{ \langle\psi|Q|\psi\rangle : \langle\psi|\psi\rangle = 1, 
\psi\in\mathcal{H}_{\infty}\}$ of $Q$ is a subset of the closed convex hull 
of the range of $Q^{b}$ in $\mathbf{C}$.

It follows that \emph{the operator norm of $Q$  is majorized  by the  
 supremum of $|Q^{B}|$ on $\mathcal{H}$}.

 \smallskip 
Let $\rho>0,\ r\in\mathbf{R}$.
\emph{A functional $F = F(\psi)$ is of the  class} 
$\mathcal{S}_{\rho}^{r}$ if 
for any  $n$ there exists a constant $C$ such that
the $n$-th Frechet differential of $F$
$$
|||(d^{n}F)(\psi)|||_{-\rho} \leq 
C(1+||\psi_{j}||_{-\rho})^{r-n},
$$
where $|||\cdot|||_{-\rho}$ is the norm of a polynomial on 
$\mathcal{H}_{-\rho}$.
 
Any such $F$ is the Berezin symbol $Q^{b}$ of a continuous linear operator
$Q:\mathcal{F}_{\infty}\rightarrow\mathcal{F}_{-\infty}$. 
The operator is a $\rho$-\emph{pseudodifferential 
operator} of the class $Op(\mathcal{S}_{\rho}^{r})$ and order $r$.

\smallskip 
The product $Q_{2}Q_{1}$ of two pseudodifferential operators 
$Q_{2}$ and $Q_{1}$ 
of orders $r_{2}$ and $r_{1}$  is a pseudodifferential operator of 
the order $r_{2}+r_{1}$.

\smallskip
Its Berezin symbol has an asymptotic expansion (cf.[14]): 
$$
(Q_{2}Q_{1})^{b}(\psi) - \sum_{n=0}^{N}\frac{(-1)^{n}}{(2n)!}
\left\langle\partial^{n}Q_{2}^{b}(\psi) | 
\partial^{n}Q_{1}^{b}(\psi)\right\rangle_{n}
\in \mathcal{S}_{\rho}^{r_{2}+r_{1}},
$$
where  $\langle\cdot | \cdot\rangle_{n}$ is the associated Hermitian 
form on the space of symmetric polynomials of order $n$, and
$\partial^{n}$ are the  complex differentials of order $n$.

\subsection{Polynomial operators.}

 For $\psi_{1}, \ldots,\psi_{k+l}\in\mathcal{H}_{-\infty} $
define continuous $kl$-\emph{monomial} operators 
$$
M^{kl}(\psi_{1}, \ldots,\psi_{k+l})
= 
\prod_{j=1}^{k}C(\psi_{j})\prod_{j=k+1}^{k+l}A(\psi_{j}):
\mathcal{F}_{\infty}\rightarrow\mathcal{F}_{-\infty}.
$$
With fixed $\Psi',\Psi''\in\mathcal{F}_{\infty}$, 
the $kl$-sesquilinear forms
$$
K^{kl}(\Psi''|\Psi') =
\langle\Psi'' | M^{kl}(\psi_{1}, \ldots,\psi_{k+l})\Psi'\rangle 
$$
are linear and symmetric in 
$\psi_{1}, \ldots,\psi_{k}$ and anti-linear and symmetric
in $\psi_{k+1}, \ldots,\psi_{k+l}$. Moreover they belong
to the core $(\overline{\mathcal{H}}^{\odot k}\otimes 
\mathcal{H}^{\odot l})_{\infty}$  of the 
Hilbert scale associated with the scaling operator 
$s^{\otimes(k+l)}$ on the Hilbert tensor 
product of symmetric Hilbert tensor powers of
$\overline{\mathcal{H}}^{\odot k}$ and $\mathcal{H}^{\odot l}$.

\smallskip
The contraction $\langle c_{kl} | K^{kl}(\Psi''|\Psi')\rangle$ 
of $K^{kl}$ with
 $c_{kl}\in(\overline{\mathcal{H}}^{\odot k}\otimes 
\mathcal{H}^{\odot l})_{-\infty}$
is a  sesquilinear  form of $\Psi',\Psi''\in\mathcal{F}_{\infty}$.
It is an integral kernel of a continuous linear operator 
$c_{kl}M^{kl}(\psi_{1},\ldots,\psi_{k+l}):
\mathcal{F}_{\infty}\rightarrow\mathcal{F}_{-\infty}$.
 
If the coefficient $c_{kl}$ is  $\rho$-\emph{continuous} with $\rho>0$, i.e., 
$c_{kl}\in(\overline{\mathcal{H}}^{\odot k}\otimes 
\mathcal{H}^{\odot l})_{\rho}$ then
$c_{kl}M^{kl}(\psi_{1},\ldots,\psi_{k+l})$ transforms  
$\mathcal{F}_{-\rho}$, to $\mathcal{F}_{\rho}$.

\smallskip
  A \emph{polynomial operator of order} $n$ with  coefficients $c_{kl}$ 
is a finite sum 
$$
P = \sum_{k+l\leq n}c_{kl}M^{kl}.
$$
\smallskip
In  $(X,\mu)$-coordinates it can be written as 
$$
P = \sum_{k+l\leq n}\int\prod_{j\leq k+l}d\mu(\chi_{j})
c_{kl}(\chi_{1},\ldots,\chi_{k+l})
\prod_{j=1}^{k}C(\chi_{j})\prod_{j=k+1}^{k+l}A(\chi_{j}).
$$
In particular, the free hamiltonian
$$
H_{0} = \int d\mu(\chi')d\mu(\chi'')\tau(\chi',\chi'')C(\chi')A(\chi'')
= \int_{X}d\mu(\chi)C(\chi)A(\chi),
$$
where $\tau$ is the trace functional on the the symmetric product
$\overline{\mathcal{H}}\odot \mathcal{H}$.

\smallskip
A polynomial operator is called $\rho$-\emph{continuous} 
if all its coefficients are $\rho$-continuous. 

\smallskip 
The coherent state matrix element a polynomial operator $P$ is 
$$
P(\psi''|\psi') = 
\sum_{k+l\leq n}c_{kl}(\psi'',\ldots,\psi'' | \psi',\ldots,\psi').
$$  
This is a holomorphic polynomial on 
$\overline{\mathcal{H}}_{\infty}\times \mathcal{H}_{\infty}.$ 

Its  Wick symbol is the real-analytic polynomial
$$
P^{w}(\psi) = \sum_{k+l\leq n}c_{kl}(\psi,\ldots,\psi | 
\psi,\ldots,\psi).
$$

\smallskip 
The Berezin symbol of a $\rho$-continuous polynomial operator $P$ can
obtained from  its Wick symbol as the \emph{finite} sum 
$$
P^{b}(\psi) = 
\sum_{j}\frac{(-1)^{j}}{(2j)!}\int D[\psi']
e^{-\langle\psi' | \psi'\rangle}
(d^{2j}P^{w})(\psi;\psi').
$$
This is well defined because the differentials
$(d^{2j}P^{w})(\psi;\psi')$  are \emph{continuous} polynomials
in $\psi'$ of order $2j$ on $\mathcal{H}_{-\rho}$ so that the 
integral can be understood as a Radon Gauss integral.

\smallskip
The principal parts of the Wick and Berezin symbols coinside:
$$
P_{0}^{w}(\psi)) = P_{0}^{b}(\psi).
$$

\smallskip
The Wick symbol of the free hamiltonian operator is 
$H_{0}^{w}(\psi)=\langle\psi | \psi\rangle$. Because it is not 
$\rho$-continuous for any $\rho>0$, 
the free hamiltonian $H_{0}$ has no Berezin symbol.

\section{Feynman equation for interacting Segal  systems.}

\subsection{Elliptic polynomial operators.}
A  polynomial operator $P\in Op(\mathcal{S}_{\rho}^{n})$  is 
\emph{elliptic} (cf.[14])  
if there are  positive constants $\rho$ and $C$ 
such that the principal Wick symbol
$$
|P_{0}^{w}(\psi)| \geq C||\psi||_{-\rho }^{n}.
$$
If, in addition, $P^{w}(\psi)$ is real
then $P$ is essentially self-adjoint on $\mathcal{F}$  
and $\mathcal{F}_{\infty}$ is its essential domain.  
 (With  the Lascar pseudodifferential calculus at hand,
the proof is similar to the finite dimensional case (cf.[18]).

Since $P_{0}^{w}= P_{0}^{b}$, the spectrum of  $P$
 is bounded from either side if 
and only if the principal Wick symbol $P_{0}^{w}(\psi)$
is bounded from the same side.  

\smallskip
A \emph{basic $\rho$-elliptic operator of order} $2$ is
$$
H_{\rho} = \int_{X}d\nu(\chi)\lambda(\chi)^{-2\rho}C(\chi)A(\chi)
$$
associated with the spectral expansion of the scaling operator $s$.

Its Wick and Berezin symbols both are equal to
$$
\int_{X}d\nu(\chi)\lambda(\chi)^{-\rho}\langle\psi(\chi) | 
\psi(\chi)\rangle = ||\psi||_{-\rho}^{2}.
$$
  
\smallskip 
 In the next theorem we represent the Wick symbol of 
 $U(t) = e^{-iPt}$ via the Berezin symbol of the generator $P$. 

\newtheorem{theorem}{Theorem}
\begin{theorem} 
Suppose $P$ is a $\rho$-continuous elliptic polynomial operator with 
the real Wick symbol $P^{w}(\psi)$.

\smallskip 
Then the coherent matrix element $\langle e^{\psi''} | e^{-iPt} |
e^{\psi'}\rangle$ of
the quantum evolution  operator $e^{-iPt}$ 
 is equal to the  limit at $N=\infty$ of the 
 functional integrals over $\mathcal{H}^{N-1}$
$$
\int\prod_{j=1}^{N-1}D[\psi_{j}]
\exp\sum_{j=1}^{N+1}\left[\langle\Delta\psi_{j} | \psi_{j}\rangle - 
iP^{b}(\psi_{j})t/N\right],
$$
where $\Delta \psi_{j} = \psi_{j}-\psi_{j-1},
\ \psi_{N+1}= \psi'',\ \psi_{0} = \psi'$.
\end{theorem}
The following  proof is adapted from [8].

First, we have the Euler-Hille limit in the strong operator topology
on $\mathcal{F}$
$$
\exp(-iPt) = \lim_{N\rightarrow\infty}(\mathbf{1} + 
iPt/N)^{-N}.
$$
Next, let $Q_{N}$ be the $\rho$-pseudodifferential operator
of order 0 with the 
Berezin symbol $Q_{N}^{b}(\psi) = [1+iP^{b}(\psi){t/N}]^{-1}$.

By the Lascar composition rule, 
$$
(\mathbf{1} + iPt/N)Q_{N} = \mathbf{1} + (t/N)^{2}R_{N},  
$$
where $R_{N}$ is a pseudodifferential operator of zero order
with the Berezin symbol bounded  uniformly relative to $N$.
By the fundamental property of the Berezin symbols,
 the norms of the operator $R_{N}$ on $\mathcal{F}$ are also uniformly
bounded. Therefore, $Q_{N}$ approximate 
$(\mathbf{1} + iPt/N)^{-1}$
with the rate $(t/N)^{2}$ in the operator norm on $\mathcal{F}$.

Thus $(Q_{N})^{N}$ strongly approximate $U(t)$ with the rate $t/N$, 
so that $(Q_{N})^{N}$ strongly converge to $U(t)$.

The coherent matrix element $\langle e^{\psi''} | (Q_{N})^{N} | 
e^{\psi'}\rangle$ is the
$(N-1)$-fold kernel contraction of the Toeplitz kernels of $Q_{N}$:

$$
\int\prod_{j=1}^{N+1}\frac{D[\psi_{j}]
\exp\langle \psi_{j}-\psi_{j-1} | \psi_{j}\rangle}
{1+iP^{b}(\psi_{j})t/N}
$$
with $\psi_{N+1}= \psi'',\  \psi_{0} = \psi'$.

Its limit at $N=\infty$ is the coherent matrix element of the evolution 
operator.

\smallskip
Replacement of  $[1+iP^{b}(\psi_{j})t/N]^{-1}$ with 
$\exp\left[-iP^{b}(\psi_{j})t/N\right]$ in the 
$(N-1)$-fold kernel contraction makes an approximation with the rate 
$N(t/N)^{2}$. 

This implies the theorem.

\smallskip
Setting 
$\psi_{j} = \psi(t_{j}),\ t_{j}=jt/N, j=0,1,2,\ldots,N$,
and  $\Delta t_{j}= t_{j+1}-t_{j}$,
rewrite the  multiple integral as
$$
\int\prod_{j=1}^{N-1}D[\psi_{j}]
\exp i\sum_{j=1}^{N}\left[-i\langle\Delta\psi_{j}/\Delta t_{j} | \psi_{j}\rangle - 
P^{b}(\psi_{j})\Delta t_{j}\right].
$$
 Its limit at $N=\infty$  is a rigorous mathematical  definition 
 of the heuristic  Feynman-Lieb integral 
$$
\int_{\psi'}^{\psi''}\prod_{0< s < t}D[\psi(s)] \exp\left [i\int_{0}^{t}d\tau
\left(-i\langle \dot{\psi}(\tau) | \psi(\tau)\rangle - 
P^{b}(\psi(\tau)\right)\right].
$$
  Theorem 1 justifies  the Feynman equation
  for $\rho$-continuous elliptic polynomial  operators $P$.
 
 \subsection{Self-interacton of the Segal system.}

A self-interaction of the  Segal system is  governed by
a Hamiltonian operator  $P$. Let us assume that $P$ is a $\rho$-continuous
 elliptic polynomial operator of 
arbitrary order with the principal Wick symbol bounded from below.

\smallskip
Let $c$ be a constant such that $P + c\mathbf{1} \geq 0$. 
Since $H_{0}$ is a  non-negative self-adjoint operator on  
$\mathcal{F}$,
$H_{0} + (P+ c\mathbf{1})$ on $\mathcal{F}_\infty$ has 
the Friedrichs self-adjoint extension  
$H_{0} \dot{+} (P+ c\mathbf{1})$ on $\mathcal{F}$ (cf.[17]). 
Denote $H = H_{0} \dot{+} P$  the self-adjoint operator
$[H_{0} \dot{+} (P+ c\mathbf{1})] - c\mathbf{1}$.  Certainly,
it does not depend on the choice of $C$.

\smallskip
The  evolution operator of the self-interacting Segal system is 
$e^{-iHt}$. 

Since $H_{0}$, and therefore $H$, is  not $\rho$-continuous, 
the theorem 1 above are not applicable to $H$ directly.

\smallskip
Consider the strong operator Trotter limit (cf.[17])
$$
e^{-iHt} = 
\lim_{N\rightarrow\infty}\left(e^{-iH_{0}t/N}e^{-iPt/N}\right)^{N}.
$$

\smallskip
Replacement of $e^{-iPt/N}$ with $(1+iPt/N)^{-1}$ does not change the 
limit (cf.[13]). 

A further replacement with $Q_{N}$ from the previous section 
preserves the limit as well:
$$
e^{-iHt} = 
\lim_{N\rightarrow\infty}\left(e^{-iH_{0}t/N}Q_{N}\right)^{N}.
$$

\smallskip
The Wick kernel of $e^{-iH_{0}t/N}$ is 
$\exp\left(e^{-it/N}\langle\psi'' | \psi'\rangle\right) $.
Its kernel contraction with the Toeplitz kernel of $Q_{N}$ can be
 approximated with the rate $(t/N)^{2}$ by an integral kernel
 $$
\mathcal{K}_{N}(\psi'',\psi') =  \exp[(i\langle\psi''-\psi' | 
\psi'\rangle -iP^{b}(\psi'))t/N].
$$
This implies 

\begin{theorem}
The coherent matrix element $\langle e^{\psi''} | e^{-iHt} |
e^{\psi'}\rangle$ of the quantum evolution  operator $e^{-iHt}$ 
 is equal to the  limit at $N=\infty$ of the 
 functional integrals over $\mathcal{H}^{N-1}$
$$
\int\prod_{j=1}^{N-1}D[\psi_{j}]
\exp\sum_{j=1}^{N}\left[\langle\Delta\psi_{j} | \psi_{j}\rangle 
-i(\langle\psi_{j+1} | \psi_{j}\rangle + P^{b}(\psi_{j}))t/N\right],
$$
where $\Delta \psi_{j} = \psi_{j}-\psi_{j-1},
\ \psi_{N}= \psi'',\ \psi_{0} = \psi'$.
\end{theorem}

\subsection{$P(\phi)$-interaction.}
 
The Wick symbol  of a $P(\phi)$-\emph{interaction  hamiltonian} 
 $P$ of degree $2n$ satisfies
 $$
 P^{w}(\psi) = P^{w}(\phi),\  \phi = (\psi + \psi^{*})/2. 
 $$
 With no dependence on complementary $\psi - \psi^{*}$, 
 such operators may not be elliptic.

\smallskip
 Nevertheless, suppose 
$P$ is a $\rho$-continuous polynomial operator of 
 order $2n$ is elliptic on $\Re^{*}\mathcal{H}$:
$$
|P_{0}^{w}(\phi)| \geq C(||\phi||_{-\rho })^{2n}, 
\phi\in\Re^{*}\mathcal{H}. 
$$
Consider the elliptic Hamiltonian $H_{\rho}\in 
Op(\mathcal{S}_{2\rho}^{2})$ from Subsection 4.1
$$
H_{\rho} = \int_{X}d\nu(\chi)\lambda(\chi)^{-2\rho}C(\chi)A(\chi).
$$
Th non-elliptic polynomial operator 
$Q = (1/2)H_{\rho}+P$ is \emph{hypoelliptic} (cf.[18]).
Indeed for any natural $m$
there is a constant $C$ such that the $m$-th Frechet differential 
$$
|||d^{m}Q^{b}(\psi)|||_{-\rho}
\leq C|Q^{b}(\psi)|
(1+||\psi||_{-\rho})^{-m},
$$
where  $|||\cdot|||_{-\rho}$ is the norm of a polynomial on 
$\mathcal{H}_{-\rho}$.

\smallskip 
Then $Q$ is essentially self-adjoint on $\mathcal{F}$ with 
 $\mathcal{F}_{\infty}$ as an essential domain.  
 (Again with  the Lascar psedodifferential calculus at hand,
the proof is similar to the finite dimensional case (cf.[18]).
Moreover, $Q$ is bounded from below because its Berezin symbol is.

On the other hand, since $\lambda(\chi)\leq 1$,
$$
|\langle \psi |(1/2)H_{2\rho}| \psi\rangle|
\leq (1/2) \langle \psi | H_{0}| \psi\rangle
$$
By the  Kato-Rellich theorem (cf.[17]), 
$H_{0}-(1/2)H_{2\rho}$ is a non-negative self-adjoint operator.
 
Finally, we have the hamiltonian 
$$
H_{0}\dot{+}P = (H_{0}-(1/2)H_{2\rho})\dot{+}Q,
$$
the Friedrichs  sum of a non-negative 
self-adjoint operator  and a bounded from below 
self-adjoint operator (cf. previous section).

\smallskip
It follows that \emph{Theorem 2 holds for  $P(\phi)$-interactions
that are elliptic on $\Re^{*}\mathcal{H}$}.

\smallskip
The theorem cannot be applied to  $\phi^{2n}$-interactions because, 
though elliptic on $\Re^{*}\mathcal{H}$, they are not 
$\rho$-continuous. Still they may be made such if cut off by the
contraction with a $\rho$-continuous coefficient.

\subsection{ Fermion  Segal systems.}

The theory of functional 
Schr\"{o}dinger equations for boson systems has a natural
 corresponding theory of fermion systems.
 
\smallskip
The Segal axioms are the same with just one exception: 

\smallskip
Instead of the Heisenberg commutation relations,
the Segal fermion system satisfies  the Clifford CCR:  
$$
 \widehat{\psi}_{1}\widehat{\psi}_{2} + \widehat{\psi}_{2}\widehat{\psi}_{1}
 = \Re\langle\psi_{1}|\psi_{2}\rangle\mathbf{1}.
$$

\smallskip
The fundamental uniqueness Segal theorem holds in the fermion case as well 
(cf.[2]). 

\smallskip
A covenient complex  Gaussian reresentations of the fermion Segal system
is in [4] (but  not in [2]). It is based on the
Grassmannian Berezin functional integral.

\smallskip
The forced boundeness of the fermionic CCR implies that in 
the Nelson-Baez theory of annihilation and creation
operators one may use the identity scaling operator.

\smallskip
 The Lascar theory of pseudodifferential operators is
valid for  fermions, provided that 
the symmetry of the bosonic formulas should be 
replaced by the sqew-symmetry of the fermionic ones.

\smallskip
The theorems of the previous section hold for fermions.
 Of course, one should use the  Grassmannian Berezin Gaussian 
 integrals. 
 
\smallskip
The further generalization to Feynman equation for
\emph{supersymmetric} Segal systems is straightforward.

\section{Appendix. Geometric Segal systems. }

In concrete applications the Segal system has 
additional features. 

\smallskip 
E.g., to distinguish between particles and their
anti-particles one may use  geometric quantization (cf.[11]).

\smallskip
Let $\overline{\mathcal{H}}$ be the anti-dual of $\mathcal{H}$.
It may be represented as the space of all anti-linear mappings
of the complex line $\mathbf{C}$ to $\mathcal{H}$. This is a complex 
Hilbert space( only the multiplication with a complex scalar is the
 multiplication with its 
complex-conjugate). In  Dirac terms $\mathcal{H}$ is the space of 
ket-vectors and $\overline{\mathcal{H}}$ is the space of the 
bra-vectors.

Actually, $\overline{\mathcal{H}}$ and $\mathcal{H}$ 
may be identified as \emph{real} vector spaces: a $\psi\in\mathcal{H}$ 
uniquely corresponds to that anti-linear mapping which takes 
$1\in\mathbf{C}$ to $\psi$. Their symplectic forms differ by the sign 
only.

\smallskip
The Hilbert sum $\overline{\mathcal{H}}\oplus\mathcal{H}$ 
is  symplectic as the sum of symplectic vector spaces.

There is a natural
complex conjugation on $\overline{\mathcal{H}}\oplus\mathcal{H}$:
$$
\overline{\overline{\psi'}\oplus\psi''} = \overline{\psi''}\oplus\psi'.
$$

\emph{Lagrangean subspaces} in a symplectic space are maximal among the 
subspaces on which the symplectic form vanishes.

A \emph{polarization} $\mathcal{L}$ of $\mathcal{H}$ is a \emph{complex 
Hilbert Lagrangean subspace} in $\overline{\mathcal{H}}\oplus\mathcal{H}$.
Its anti-dual is identified with the 
complex-conjugate $\overline{\mathcal{L}}$.

\smallskip
A polarization $\mathcal{L}$ is \emph{real} if  
 $ \mathcal{L} = \overline{\mathcal{L}}\cap\mathcal{L}$.
Real polarizations are the complexifications of \emph{real} Hilbert
Lagrangean subspaces in $\mathcal{H}$.

\smallskip
Let $\sigma$ be the symplectic form on 
$\overline{\mathcal{H}}\oplus\mathcal{H}$.
A polarization $\mathcal{L}$ is \emph{non-negative} 
if the quadratic form $\beta(\psi',\psi'') = i\sigma(\overline{\psi'},\psi'')
$ is non-negative on $\mathcal{L}$. Real polarizations are non-negative. 

If the quadratic form $\beta$ is positive definite on a polarization, 
then we have a \emph{K\"{a}hler polarization} of $\mathcal{H}$.

\smallskip
The non-negative polarizations 
$\mathcal{L}$ of  $\mathcal{H}$ define  representations of the Segal 
system over $\mathcal{H}$. We call such representations the 
\emph{geometric Segal systems}.

The real Gaussian representation is the geometric Segal system
 corresponding to 
the real diagonal polarization $\{\overline{\psi}\oplus\psi:
\psi\in\mathcal{H}\}$.

The complex Gaussian  representation is the geometric Segal system 
corresponding to 
the K\"{a}hler polarization $\mathcal{H} = 0\oplus\mathcal{H}$. 

\smallskip
In physics terms, the geometric Segal systems which correspond 
to anti-dual polarizations describe a pair of free anti-particles.
 One may think about the complex conjugation as a 
generalization of the CPT transformation.

\smallskip 
Geometric Segal systems are \emph{geometrically equivalent}
if the corresponding polarizations are equivalent 
under  transformations of  
 $\overline{\mathcal{H}}\oplus\mathcal{H}$ induced by  
  symplectic  automorphisms of $\mathcal{H}$.
 
\emph{Two non-negative polarizations are geometrically equivalent  if 
 and only if the nullity of the restriction of the quadratic form $\beta$
is the same for both polarizations.}

\section{ How constructive quantum field theory is possible.}

There are three basic formulations of constructive quantum field theory
(cf.[6]).

\smallskip
1. \emph{Canonical formulation.}

Quantum fields are operator-valued fields on the Minkowski space-time 
$\mathbf{R}^{d,1}$ that 
 satisfy the canonical commutation relations
and solve the  classical  Hamiltonian equations.
For interacting fields the equations are \emph{non-linear}
 partial differential equations on $\mathbf{R}^{d,1}$. 

\smallskip
Unfortunately for $d > 3$, 
the relativistic irreducible quantum fields, 
which satisfy the canonical commutation relations, are  free by default
(cf.[3]).

Even the $d=3$ case is  troublesome. The simplest non-linearity in the
 perturbation theory is the $\phi^{4}$ interaction.
  Yet for $d = 3$  renormalization screens out the perturbation 
  (cf.[7]).

\smallskip
2. \emph{Feynman formulation.}

The quantum propagators of classical fields are  Feynman integrals 
over classical  histories on the Minkowski space-time $\mathbf{R}^{d,1}$.

\smallskip
Since 1960's the prevalent approach is the Lagrangean Feynman-Kac   
 infinite-dimensional  integral
over the space of histories on the \emph{euclidean} space 
$\mathbf{R}^{d+1}$ with the \emph{aposteriory} 
analytic continuation to the  real time. 

This approach of K.Symanszik 
and E.Nelson has culminated in the work of J.Glimm and A.Jaffe [10].
However, its application to interacting fields
 in the space dimensions  $d > 2$ is still open.

3. \emph{Functional Schr\"{o}dinger formulation.}

The quantum states are functionals on the phase space propagated by 
the evolution operator of a \emph{linear} functional differential
Schr\"{o}dinger equation.

\smallskip
For quite some time the functional Schr\"{o}dinger formulation 
has been presumed  mathematically unreasonable.
To quote F.Berezin [4]: 
\begin{quote}
\ldots the mathematical problems 
occuring in the method of second quantization are somewhat removed 
from the problems of the traditional mathematical physics which 
are formulated in 
terms of partial differential equations. In particular,  major roles 
in the  method of second quantization are taken by purely algebraic 
questions, strange to classical mathematical physics\ldots

\end{quote}
Yet important analytic techniques for  functional differential
 equations have 
been developed in the P. Kr\'{e}e seminar at the Institut Henri
Poincar\'{e} in Paris during 70's. The 
B.Lascar theory [14] of  ifinite-dimensional 
pseudodifferential operators is its byproduct.

\smallskip
We have solved the \emph{linear} functional Schr\"{o}dinger equations 
for  interacting  Segal boson and fermion  
systems with $\rho$-continuous (hypo) elliptic polynomial interactions. 
The results are applicable to quantum fields on Minkowski space-times
 of \emph{any} dimension.

The solutions are Feynman type sequential integrals defined as
limits of multiple integrals over the phase space. 
Seemingly unpractical for concrete  computations, 
they justify the basic integral calculus rules. 
In this respect \emph{the approximating multiple integrals play the role of the
Riemann integral sums in elementary integral calculus providing a 
foundation for more sofisticated techniques}.

\smallskip
Unfortunately,  gauge fields do not admit elliptic interactions. 
However, presumably our approach is sound even in this case (the work 
is in progress).

\bigskip
\Large 
\textbf{References.}

\bigskip
\large
1.\ Baez,J.C.: Wick products of the free Bose field, 
  \emph{J. Funct. Anal.}\textbf{86} (1989), 211--225. 

\smallskip
2.\ Baez,J.C., Segal,I.E.,and Zhou,Z.-F.:
 \emph{Introduction to algebraic and constructive
quantum field theory}, Princeton University Press, Princeton, 1992. 

\smallskip
3.\ Baumann,K.: On relativistic irreducible quantum 
fields fulfilling CCR,\emph{J. Math. Phys.} 
 \textbf{28}(1988), 697-704.

\smallskip
4.\ Berezin,F.A.: \emph{The Method of Second 
Quantization}, New York, Academic Press, 1966.

\smallskip
5.\ Bogoliubov,N.N., Logunov,A.A., Oksak,A.I., Todorov,I.T.:
 \emph{General Principles of Quantum Field Theory}, Kluwer, 
 Dordrecht, 1975.

\smallskip
6.\ Bogoliubov,N.N., Shirkov,D.V.:
 \emph{Introduction to the theory of quantized fields},
 Wiley, New York, 1980.

\smallskip
7.\ Callaway,D.J.E.: Triviality pursuit: can elementary
scalar particles exist?, \emph{Phys.Reports} \textbf{167 }(1988) 241-320. 

\smallskip
8.\ Dynin,A.: A rigorous path integral construction in any 
dimension, \emph{Letters Math. Phys.},\textbf{34} (1998), 317-329.

\smallskip
9.\ Feynman,R.P.,An operator calculus having applications 
in quantum electrodynamics, \emph{Physical Review} \textbf{84} 
(1951), 108-128.

\smallskip
10.\ Glimm,J., Jaffe,A.: \emph{Quantum physics:
 a functional integral point of view}, Springer-Verlag, New York, 1981.

\smallskip
11.\ Kirillov,A.A.: Geometric quantization,
 in \emph{Dynamical systems IV},
Arnol'd,V.I.,Novikov, S.P.(eds.), Springer-Verlag, New York,1990.

\smallskip
12.\ Klauder,J.R.: The action option and a Feynman 
quantization of spinor fields in terms of ordinary c-numbers,
\emph{Ann.Phys.(N.Y.)} \textbf{11} (1960),123-168.

\smallskip
13.\ Lapidus,M.L.: Product formula for imaginary resolvents 
with application to a modified Feynman integral,
\emph{Journal of Functional Analysis} \textbf{63} (1985), 261-275.

\smallskip
14.\ Lascar,B.: Une condition necessaire et suffisant 
d'ellipticite pour une class d'operateurs differentiells en dimension 
infinie, \emph{Comm. P.D.E.} \textbf{2} (1977), 31-67.

\smallskip
15.\ Lieb,E.H.: The classical limit of quantum spin systems,
\emph{Comm. Math. Phys.} \textbf{77} (1973), 127-136.

\smallskip
16.\ Nelson,E.: Time-ordered operator products of sharp-time 
quadratic forms, \emph{J. Funct. Anal.} \textbf{11} (1972), 211--219.

\smallskip
17.\ Reed,M., Simon,B.: \emph{Methods of modern mathematical 
physics}, II, Academic Press , New York, 1972.

\smallskip
18.\ Shubin,M.A.: \emph{Pseudodifferential operators and spectral 
theory}, Springer-Verlag, New York, 1987.  

\smallskip
19.\ Tobocman,W.: Transition amplitudes as sums over 
histories, \emph{Nuovo Cim.} \textbf{3} (1956), 1213-1229.

\end{document}